\documentclass[preprint,showpacs,showkeys,aps]{revtex4}
\usepackage{latexsym}
\usepackage{graphicx}

\begin{document}

\title{Topology of Foreign Exchange Markets using Hierarchical Structure Methods}

\author{Michael J. Naylor$^{(1)}$, Lawrence C. Rose$^{(2)}$, and Brendan J. Moyle$^{(2)}$}

\affiliation{\small{$^{(1)}$ Corresponding Author, Department of Finance, Banking and Property, Massey University, New Zealand, M.J.Naylor@massey.ac.nz\\
$^{(2)}$Department of Commerce, Massey University}}

\begin{abstract}

This paper uses two physics derived hierarchical techniques, a minimal spanning tree and an ultrametric hierarchical tree, to extract a topological influence map for major currencies from the ultrametric distance matrix for 1995 - 2001. We find that these two techniques generate a defined and robust scale free network with meaningful taxonomy. The topology is shown to be robust with respect to method, to time horizon and is stable during market crises. This topology, appropriately used, gives a useful guide to determining the underlying economic or regional causal relationships for individual currencies and to understanding the dynamics of exchange rate price determination as part of a complex network.

\pacs{\textit{02.50.Sk, 89.65.-s, 89.65.Gh, 89.75.Hc}}

\keywords{\textit{minimal spanning tree, ultrametric hierarchical tree, taxonomy, econophysics, financial markets}}
\end{abstract}

\maketitle
 
\section{\textbf{Introduction}}

Hierarchical structure methods are used in finance to ascertain the 
structure of asset price influences within a market. These methods use the 
synchronous correlation coefficient matrix of daily difference of log prices 
to quantify the pricing distance between assets in terms of the inherent 
hierarchical structure. This structure will give some indication of the 
taxonomy of an assets' portfolio, and can be used to generate an asset 
markets' hierarchy. 

Two techniques will be used in this paper. The first technique is the 
creation of a\textit{ minimal spanning tree} (MST), which is a graph of a set of $n$ elements of the 
arrangement of the nodes in an \textit{ultrametric space}. MST has been shown to provide sound results 
for financial assets with the resultant taxonomy displaying meaningful 
clusters [1, 2, 3, 4]. MST also helps to overcome the empirical problem of 
noise in a historical correlation matrix [5]. 

The second technique is the creation of an \textit{ultrametric hierarchical tree} structure [6, 7]. This technique 
gives a determination of the hierarchical structure of a network and is 
particularly useful for determining if hubs exist.

The structure of asset price movements is extracted by use of a synchronous 
correlation coefficient matrix, $A_{ij}$, of daily difference of log prices. 
This matrix is transformed [8] by the equation below to get the ultrametric 
pricing distance between currencies. This metric preferred to correlation as 
it fulfils the three axioms of a metric distance [1].
\[
d(i,j)=\sqrt {2(1-A_{ij} )} 
\]
The choice of clustering procedure is vital as it has more effect on the 
quality of clustering than the choice of distance metric [9]. MST analysis 
uses the single-linkage clustering method which builds up clusters by 
starting with distinct objects and linking them based on similarity. The 
major issue with this method is that while it is robust for strongly 
clustered networks, it has a tendency to link poorly clustered groups into 
chains by successively joining them to their nearest neighbours [10]. These 
chains are non-robust to data variation, and thus MST is less robust for 
larger distances. The information obtained should thus be used with care and 
be combined with other techniques if possible. This paper will focus on the 
extraction of price influences rather than on determinants of market 
activity.

\section{\textbf{The data}}

Forty-four currencies (table 1) were chosen because they were generally free 
floating, covered the data period (23/10/95 - 31/12/01) and had either 
market dominance or represented a region. The Mexican peso and Russian 
rouble were used in their format prior to currency reforms, which removed 
three zeros. Data were sourced from Oanda.com at Olsen and Associates. The 
exchange rates were daily average inter-bank ask rates as determined in 
Zurich. This should give some idea of how international currencies interact, 
how the currency nodes are clustered, and the pattern behind price 
influences. This is a small sample compared to stock market studies, which 
will limit possible topologies.

\section{\textbf{Numeraire}}

One of the problems uniquely encountered in foreign exchange research is 
that currencies are priced against each other so no independent numeraire 
exists. Any currency chosen as a numeraire will be excluded from the 
results, yet its inherent patterns can indirectly influence overall 
patterns. There is no standard solution to this issue or a standard 
numeraire candidate. Gold was considered, but rejected due to its high 
volatility. 

This is an important problem as different numeraires will give different 
results if strong multidimensional cross-correlations are present. Different 
bases can also generate different tree structures. The inclusion or 
exclusion of currencies from the sample can also give different results. 
This implies samples should include all major currencies and undue emphasis 
should not be placed on any particular MST result. Result robustness should 
also be checked by comparison with other methods or samples. 

This study used both the NZD and the USD as numeraires. The NZD is a minor 
currency which can be easily excluded, and it does not impose any strong 
default pattern. However the overwhelming dominance of the US dollar tends 
to submerge secondary influences. Use of the US dollar as the numeraire 
allows second-order relationships to be examined, as the MST will show 
differences in price influence. The results are indicative only however as 
the exclusion of the US dollar can impose a default pattern due to 
cross-correlations. A larger sample size was used for the US dollar tests to 
allow regional clusters to be developed. 

An alternative approach to MST graphs is the use of all currency pairs [3]. 
This is a valid approach but it does add more complexity, gives visual 
results which are difficult to interpret, as well as potentially missing out 
influential currencies. The approach is also impractical if additional 
causal links in addition to the primary link are examined, or if the sample 
size is larger than ten. There are also problems caused by the impact on 
correlations of cross-quotations, as this imposes a default structure. 

\section{\textbf{Results from total period NZD matrix}}

The NZD based distance-metric matrix indicated similarity in currency 
dynamics between NLG-BEF (0.1743), NLG-FRF (0.1805) and BEF-FRF (0.2125), 
all members of the European Monetary System (EMS). The main surprise was the 
DEM which is not close to any of the EMS countries, but it is reasonably 
close to the MYR (0.5191). MYR-SGD (0.6888) was the only close currency pair 
in SE Asia (SEA), though inter-Asian values tend to be lower than 
intra-Asian. 

The minimum spanning tree (MST), shown in Figure 1, was created using 
Pajek$^{\copyright }$ [11] and Kruskal's algorithm. The star like structure 
indicates the USD is the predominant world currency, and the hub of an 
international cluster with only the EMS cluster separate. Inter-SEA FX 
linkages are stronger than in other (non-EMS) regions, with the IDR linked 
via the MYR, and THB and PHP linked to the SGD. The GBP links to the USD in 
preference to the EMS and the AUD links via its commodity cousin, the CAD. 
Two other commodity currencies are also linked, the BRL and the CLP.

\begin{figure}[htbp]
\centerline{\includegraphics[width=3.85in,height=2.90in]{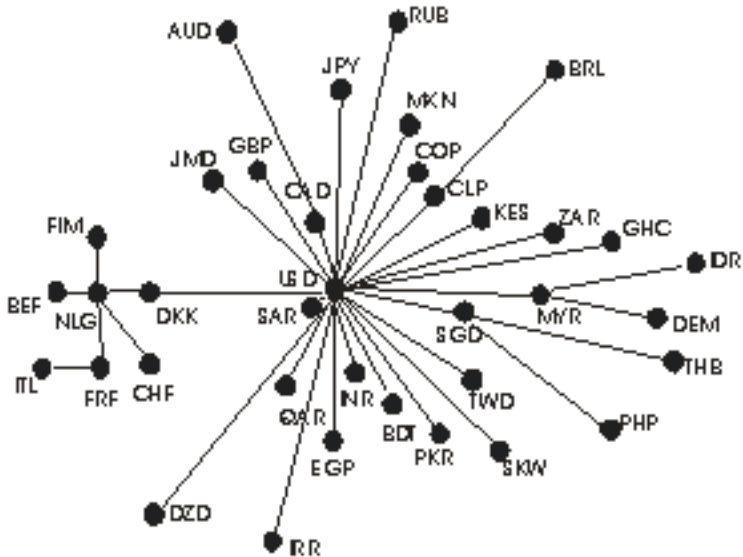}}
\caption{Fig 1 -- NZD-based FX minimum spanning tree (1995-2001). This gives a graphical representation of minimal distance metrics for currencies quoted against the NZD.This gives an indication of the basic first-order price causation determination. The USD is shown as the hub, with an attached EMS cluster.} 
\label{fig1}
\end{figure}

The linking of the DEM to the MYR is a puzzle as the mechanism of the EMS 
implies close correlations between EMS members. This implication is backed 
by correlation based empirical evidence [12, 13, 14]. One possible 
explanation is that as the lead currency within the EMS the value of DEM is 
determined by global influences and these are best reflected in the value of 
the MYR. Other EMS currencies are followers so do not respond on a daily 
basis to global factors. Another explanation is that since the distance 
metric is based on log daily changes with no lag component it reflects 
short-term movements within the EMS band. For the DEM these may not be 
related to the intra-band movements of other EMS currencies. This issue 
needs further analysis. The DEM result emphases the causation made in 
section 1; that hierarchical structure techniques need to be used with care 
and combined with other techniques if possible. 

MST analysis shows the key determinant European currency as the NLG with the 
inter-cluster linkage via the USD-DKK. It is of interest that currencies 
which are isolated, like the RUB or the IRR, still have the USD as their 
main determinant link. The MST was robust to excluded currencies, with these 
either being USD linked (CRC) or EMS linked (CZK). Further studies using 
monthly and annual data indicated no change in the basic topology.

The distribution of links per node is more centralised than a power-law 
would dictate. An ln-ln plot of link density had a slope of 0.8, indicating 
a strong element of self-organisation in the international financial system. 
These results imply that either currencies are all linked by common economic 
factors or currency traders pay more attention to USD movements than to 
local factors. The spread of most distance values in the 0.8 to 1.16 range 
also reinforces the point that price setting in currency markets is 
generally more homogeneous than price setting in stock markets. 

Our results show a more centralised arrangement than US stock market 
studies. The DJ and the S{\&}P 500 indexes (1989-1995) had a closest 
distance of 0.949, with most in the 1.09 to 1.3 range [1]. Stock indices 
were also more web-like structure with four distinct clusters for the DJIA, 
16 major clusters for the S{\&}P with 44 minor clusters, and 18 key stocks 
acting as linkages between S{\&}P clusters. Similar results topology was 
found for stocks on the Nasdaq, NYSE and AMEX [2, 15], with a power-law with 
degree 2.2 for the distribution of links per node, and the non-random nature 
of the MST remained remarkably consistent over time. 

While the small size of our sample obviously can not be expected to generate 
complex arrangements, the simplicity is mainly due to the pre-dominance of 
the USD with currency markets. This is a well-established empirical result 
[16], which has not been found to exist in the stock market. The simplicity 
of our results was also robust over the entire set of 208 currencies. Onnela 
[18] did find a similar star-based network for the S{\&}P 500 index, though 
the methodology has been criticised by [19].

The hierarchical tree of the subdominant ultrametric space associated with 
the MST is shown in Figure 2.

\begin{figure}[htbp]
\centerline{\includegraphics[width=5.62in,height=2.74in]{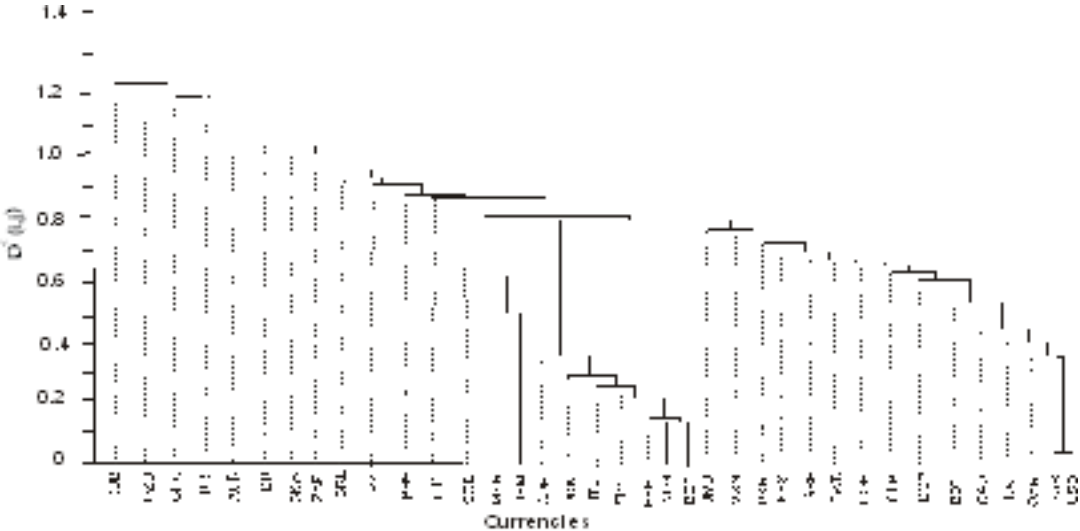}}
\caption{Fig 2 -- NZD-based FX hierarchical tree of subdominant ultrametric 
space (1995-2001).Hierarchical grouping of distances metrics for currencies quoted against the 
NZD. This gives clusters, based on primary causal link. This shows that 
price determination leads off the USD with Asian and EMS clusters.}
\label{fig2}
\end{figure}

The smoothness of the hierarchical tree shows the dominance of the USD, as 
all currencies, aside from the EMS cluster and the DEM-MYR-SGD triage, link 
off the main tree. The large distances involved for IRR, GHC, DZD and RUB 
show them to be largely isolated.

\section{\textbf{Results from total period USD matrix}}

The USD based distance-metric matrix showed weaker links than for the NZD 
matrix, with a number of links close to the no-relationship value. This is 
expected as these are secondary influences, and some currencies may not have 
influential secondary linkages. Note that care is needed when interpreting 
second order trees as relationships can be created indirectly by a joint 
association to the missing hub, the USD, rather than a direct relationship. 
Results should thus be checked against economic reasons or against other 
samples or methods.

The USD based minimum spanning tree, Figure 3, shows more clustering than 
the NZD MST star diagram, with groups nested within other groups. Visually 
relationships overall seem weaker, though the EMS hub still exists. This hub 
is stable as dropping out the USD does not affect price causation. The 
removal of the USD has, however, affected most other currency relationships. 
The GBP is now showing its second order link to the EMS as is the DZD. There 
is also a minor Scandinavian grouping. The distribution of node links of the 
USD MST is more varied than the NZD MST, with an approximate power-law 
distribution of degree 1.5 ($R^{2}$ = 0.95). Further studies using annual 
data indicated no change in the basic topology.

The MST has a number of dangling pendants. While these need to be treated 
with caution, as discussed, there does tend to be economic causation behind 
most of the linkages. The strongest economic clusters are the two commodity 
clusters. The first is the AUD, CAD, ZAR, SAR cluster. The AUD is 
conventionally used in currency markets as a play on commodity prices. The 
linking of the commodity economies to the AUD lends support to this 
convention. 

\begin{figure}[htbp]
\centerline{\includegraphics[width=4.68in,height=3.01in]{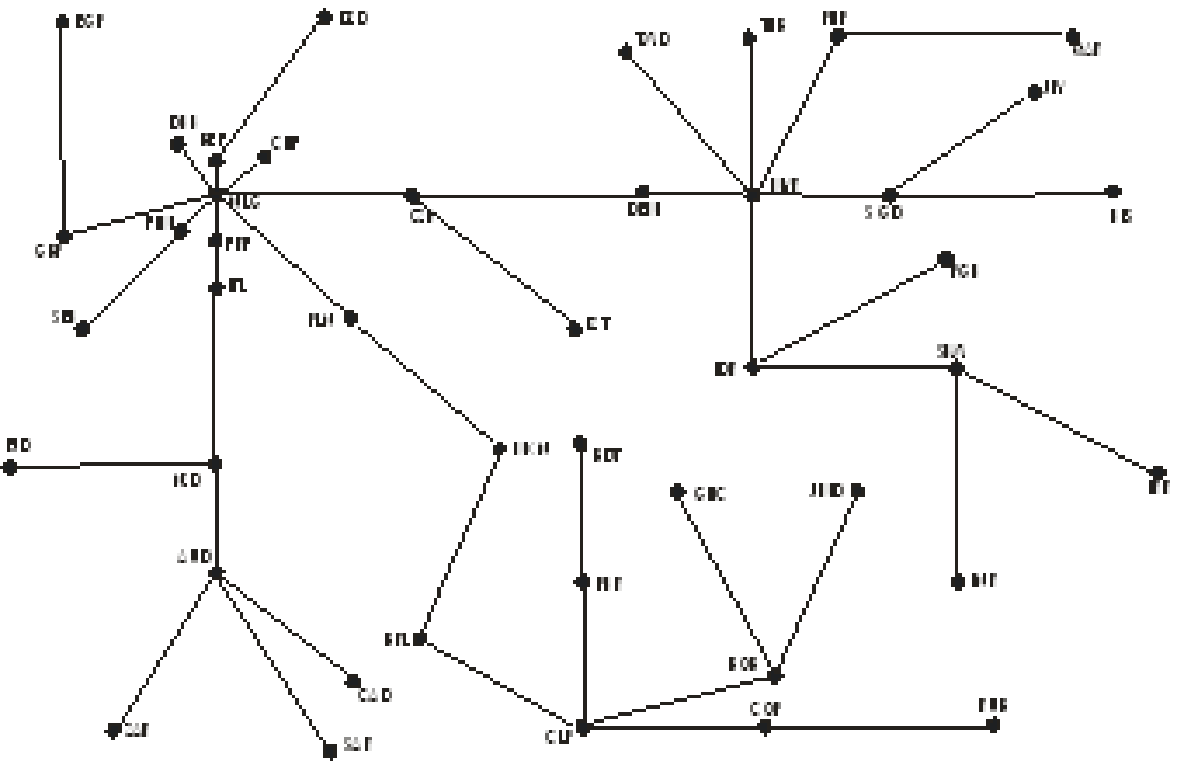}}
\caption{Fig 3 -- USD-based FX minimum spanning tree (1995-2001). Graphical representation of minimal distance metrics for currencies quoted against the USD. This gives an indication of thesecond order price causation determination. This shows a sparse clustering compared to Figure 1.}
\label{fig3}
\end{figure}

Another commodity cluster is grouped around the CLP, COP, RUB, BOB, GHC and 
JMD. The Indian subcontinent currencies of PKR and BDT are also weakly 
linked. The MXN-PLN link of this cluster back to the EMS involves long 
distances so maybe spurious. The distance involved with the RUB link 
indicates it is also isolated. The BOB, PKR and BDT links are problematic as 
they were pegged to the USD for part of the data period and have thin 
markets. 

The link between the EMS cluster and the rest of the world is via the 
NLG-CZK-DEM-MYR link, which provides the backbone to the system. A SE Asian 
cluster is evident, centred on the MYR and linked via the DEM. The SGD seems 
to be linked externally instead of inter-Asean as illustrated by its JPY 
link. The SGD/ KES link and the PHP/QAR link are probably spurious. 

The associated hierarchical tree of the sub-dominant ultrametric space, 
Figure 4, shows three main secondary clusters, the EMS, with low distances, 
the North/SE Asian, with medium distances, and the Latin American/South 
Asian grouping. Outside those clusters distances tend to be high, indicating 
ties are weak. The tree indicates the AUD based commodity cluster is part of 
the dominant USD/EMS hierarchy rather than an isolated cluster. Conversely 
the hierarchical tree indicates that the Latin American-South Asian cluster 
is isolated from the main tree, and has a separate price determinate 
process, though inter-region linkages are weak. There is no separate 
South-Asia cluster.

\begin{figure}[htbp]
\centerline{\includegraphics[width=5.95in,height=2.38in]{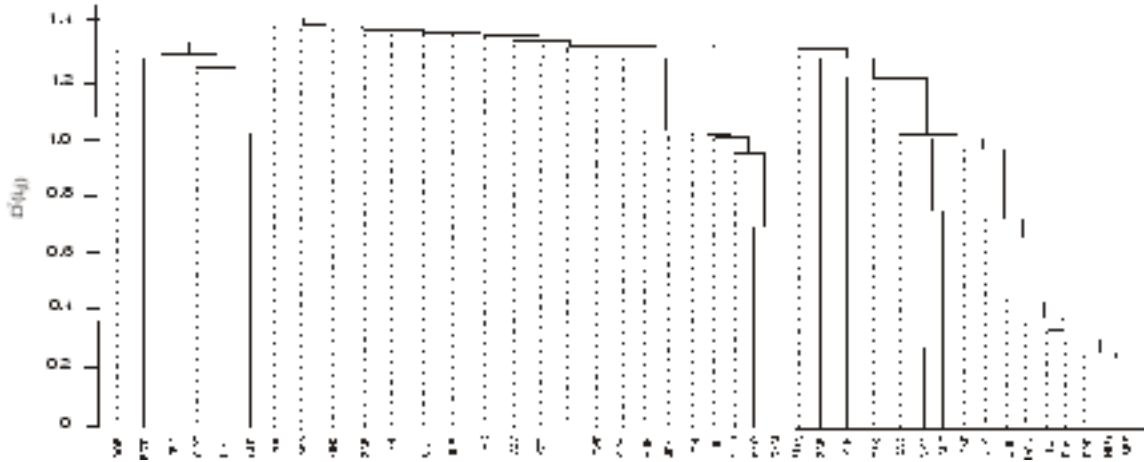}}
\caption{Fig 4 -- USD-based FX hierarchical tree of subdominant ultrametric 
space (1995-2001). Hierarchical grouping of distances metrics for currencies quoted against the 
USD. This gives an indication of how currencies should be grouped into 
clusters, based on secondary causal links. This shows EMS, Asian, and Latin 
American clusters.}
\label{fig4}
\end{figure}

\section{\textbf{Asian crisis period (1997/98)}}

The techniques were repeated for the Asian crisis period, 1$^{st}$ August 
1997 to 31$^{st}$ October 1998. The rationale is that several empirical 
studies [19] have indicated that causal determination behind currency 
movements differs during crisis periods and non-crisis periods with regional 
correlations tending towards unity. It is useful to verify these results 
with MST analysis, as this may aid our understanding of cluster dynamics 
during market crises. The results can also be compared to the stick market 
topological crisis studies which show a universal shortening of distance [5, 
17, 20].

Both the NZD and the USD distance-metric matrices show correlations within 
SE Asia increased during the crisis period, in most cases by 50 to 100{\%}, 
approximating inter-EMS correlations. North Asian currencies had lower 
distances to SE Asian currencies during the crisis period. Several non-Asian 
currencies BDT, BRL, RUB and GBP, also became more strongly linked to the SE 
Asian currencies during the crisis. In contrast to these results inter-EMS 
distances increased, indicating stress. These results may imply that 
currency traders started to treat the crisis countries as a distinct bloc 
during the period of the crisis.

While these results provide support for the hypothesis that crisis affected 
countries formed a closely tied cluster, these conclusions need to be 
treated with care as the decrease in distances of all the affected countries 
to the USD raises the alternative hypothesis of an increase in power of the 
USD in price setting. The crisis-affected countries could thus be seen as 
individually increasing their joint co-movement against the USD, and only 
indirectly moving together.

These changes are illustrated in the crisis period NZD-based minimum 
spanning tree, shown in Figure 5. The total period two-cluster network shown 
in Figure 1 is still retained during the crisis period. The EMS cluster is 
largely unchanged, though the distances are increased and the GBP and JPY 
have been picked up. The USD cluster structure is also largely unchanged 
though some relative distances have changed. In particular the RUB is now 
closer to the USD. The Asian offshoot is retained but now the THB is the key 
currency with MXN and SKW now linked. This indicates the effect of the 
crisis on those currencies.

The similarity of Figure 5 to Figure 1 indicates that channels for crisis 
propagation were activities within the USD hub impacting on one country 
after another, instead of one currency directly affecting another. These 
conclusions are strengthened by examination of the hierarchical tree of the 
subdominant ultrametric space, shown in Figure 6. This tree shows despite 
decreased distances within Asia and increased EMS distances, there are still 
only two dominant clusters. The pattern of network clustering is unaltered. 
The only noticeable change is that previously isolated currencies, like the 
RUB, IRR or GHC, are now integrated.

\begin{figure}[htbp]
\centerline{\includegraphics[width=5.20in,height=2.61in]{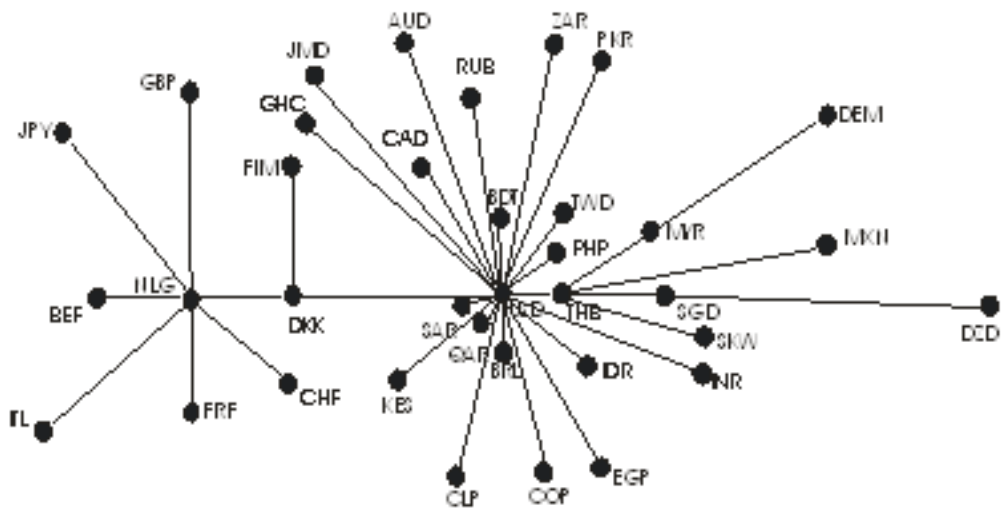}}
\caption{Figure 5 - Crisis Period FX NZD-based minimum spanning tree 
(1997-98). Graphical representation of minimal distance metrics for crisis period of 
currencies quote against the NZD. Minimal changes have occurred compared to 
Figure 1, though lengths are shorter.}
\label{fig5}
\end{figure}

\begin{figure}[htbp]
\centerline{\includegraphics[width=5.80in,height=2.85in]{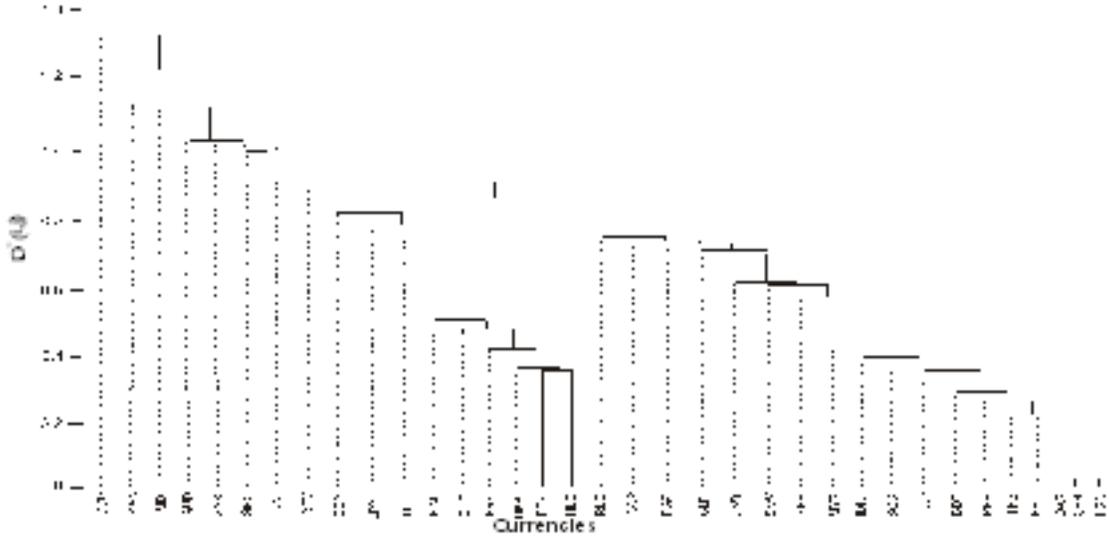}}
\caption{Figure 6 - Crisis Period FX NZD-based hierarchical tree (1997-98). Hierarchical grouping of distances metrics for crisis period of currencies quoted against the NZD. The main difference from Figure 3 is shorter distances.}
\label{fig6}
\end{figure}

The crisis period USD MST, Figure 7, shows the basic structure is retained 
to the normal period MST in Figure 3, with a defined EMS cluster, an Asian 
cluster, and a developing country commodity cluster. The primary core of 
NLG-CZK-SGD-MYR-DEM still exists. However there are some changes in the 
secondary influence pattern, as the Latin American-developing country 
commodity branch is broken up, the SGD is more central and currencies with 
more distant links have randomly re-arranged themselves. The distribution of 
node links has a slightly more even distribution, with a power-law 
distribution of degree 1. 

\begin{figure}[htbp]
\centerline{\includegraphics[width=4.64in,height=3.19in]{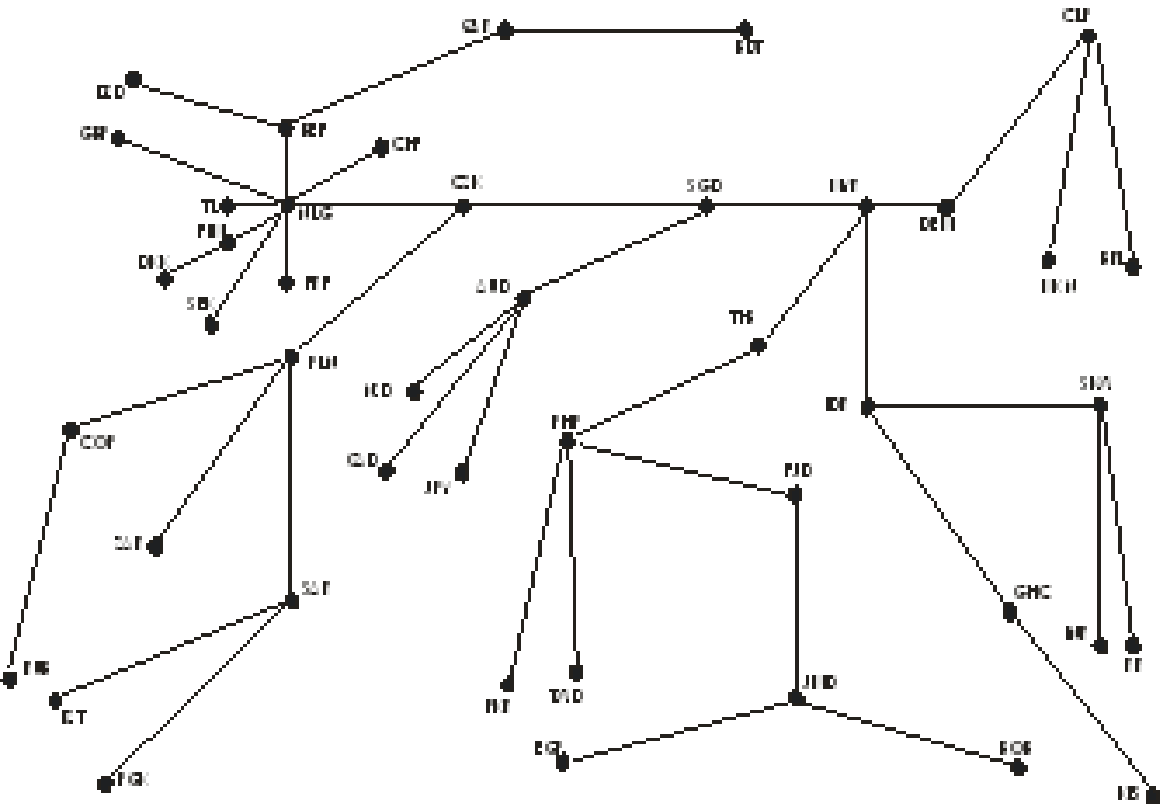}}
\caption{Figure 7 -- Crisis Period FX USD-based minimum spanning tree 
(1997-98). Graphical representation of minimal distance metrics for crisis period of 
currencies quoted against the USD. The major difference from Figure 3 is the 
increased centrality of the Asian cluster.}
\label{fig7}
\end{figure}

These conclusions are reinforced by the hierarchical tree of the subdominant 
ultrametric space Figure 8. The EMS cluster is still present and the 
Asian-based cluster is weakened, with only the DEM-MYR-SGD triage present. 
Other Asian currencies tend to co-move with this triage only as part of a 
global currency co-movement. Most non-EMS-Asian currencies have weak 
distances. The Latin-American/ South Asian cluster has disappeared. The 
overall impression is of less regional clustering during the crisis and more 
of a common global cluster, especially outside the EMS.

\begin{figure}[htbp]
\centerline{\includegraphics[width=5.72in,height=2.50in]{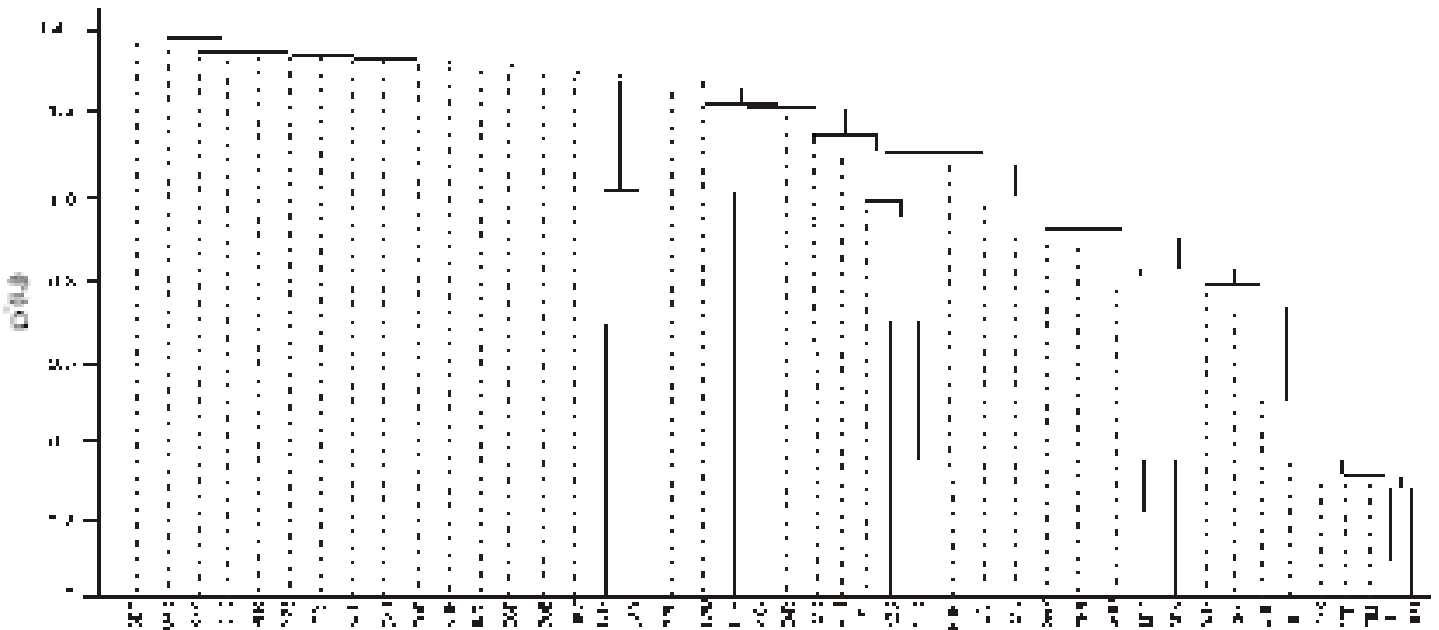}}
\caption{Figure 8 - Crisis Period FX USD-based hierarchical tree (1997-98). Hierarchical grouping of distances metrics for crisis period of currencies quoted against the USD. When compared to Figure 5 fewer clusters are evident.}
\label{fig8}
\end{figure}

\section{\textbf{Conclusions}}

This paper has shown that hierarchical methods can be used to analysis 
foreign exchange price influences. The results show price determination in 
international currency markets displays sparse clustering. The network has a 
simple tree-like structure with a dominant spine. Underneath the predominant 
influence of the USD and the EMS, there are clear secondary relationships 
based on economic or regional factors. This topology was shown to be robust 
to time horizon and market crises. The paper also indicates that the 
transmission process for cascading shocks is primarily through the spine and 
then through links outside of that spine. 

The anomalous placement of the DEM within the Asian cluster however 
indicates that care is needed with regard to interpretation of results. This 
ambiguity, together with the known tendency of the single-linkage clustering 
method to generate unstable spurious linkages, means that MST results should 
always be handled with care, and used together with other methods when 
analysing network structure. 

Overall the results also provide an indication that the price determination 
structure of international currency markets is tree like and sparsely 
clustered. This implies dynamic behaviour related to complex networks can be 
applied to currency markets. 

\section{References:}

[1] Mantegna, R. N (1999), `Hierarchical Structure in Financial Markets', 
Eur Phys Jou B: Cond Mat, Vol 11: 193

[2] Bonanno, G., N. Vandewalle {\&} R. N. Mantegna (2000), `Taxonomy of 
Stock Market Indices', Phys Rev E, Vol 62, No 6: R7615-7618.

[3] McDonald, M., O. Suleman, S. Williams, S. Howison {\&} N. F. Johnson 
(2005) `Detecting a Currency's Dominance or Dependence using Foreign 
Exchange Network Trees', Phys Rev E, Vol 72: 046106.

[4] Bonanno, G., F. Lillo {\&} R. N. Mantegna (2001) `High Frequency 
Cross-correlation in a Set of Stocks', Quan Fin, Vol 1: 96-104.

[5] Onnela, J-P., A. Chakraborti, K. Kaski {\&} J. Kert\'{e}sz (2003) 
`Dynamic Asset Trees and Black Monday', Phys A, Vol 324: 247.

[6] Laloux, L, P. Cizeau, J-P. Bouchard {\&} M. Potters (1999) `Noise 
Dressing of Financial Correlation Matrices', Phys Rev Lets, Vol 83: 1467 

[7] Plerou, V, P. Gopikrishnan, B. Rosenow, L.A.N. Amaral {\&} H.E. Stanley, 
(1999) `Universal and Non-universal Properties of Cross Correlations in 
Financial Time Series,' Phys Rev Lets, Vol 83: pp 1471 

[8] Gower, J. C. (1966) 'Some distance properties of latent root and vector 
methods used in multivariate analysis', Biometrika, Vol. 53, (3/4): 325-33

[9] Hirst, P (2003) `Cluster Analysis of Financial Securities', M.Sc. 
Thesis, Oxford Centre for Industrial and Applied Mathematics, Oxford 
University.

[10] Kaufman, L. {\&} P.J. Rousseeuw (1990) `Finding Groups in Data: An 
Introduction to Cluster Analysis', Wiley-Interscience; New York, USA. 

[11] Baragelj, V. {\&} A. Mrvar (2005) `Pajek -- Program for Large Network 
Analysis'.

[12] Parikh, A {\&} R. Bhattacharya (1996) 'Exchange Rates under EMS target 
zones: an econometric investigation', Apld Econs, Vol 28(4): 453-466.

[13] Laopodis, N.T. 'Stochastic Behaviour of Deutsche Mark Exchange Rates 
within EMS', Apled Fin Econs, Vol 13(9): 665-676.

[14] Von Hagen, J {\&} M. Fartianni (1990) 'German Dominance in the EMS: 
Evidence from Interest Rates', J. Int Money, Vol 9: 358-375 

[15] Vandewalle, N, F Brisbois {\&} X Tordoir (2001) `Non-Random Topology of 
Stock Markets', Quant Fin, Vol 1: 372-374.

[16] Frankel, J., G. Galli {\&} A. Giovannini (1996) 'The Microstructure of 
Foreign Exchange Markets', Chicago; University of Chicago Press. 

[17] Ara\'{u}jo, T {\&} F. Louc\~{a} (2005) `The Geometry of Crashes - A 
Measure of the Dynamics of Stock Market Crises,' arXiv.org/physics/0506137

[18] Onnela, J-P. (2002) `Taxonomy of Financial Assets,' Masters Thesis, 
Laboratory of Computational Engineering, Helsinki University of Technology.

[19] Johnson, N. F., M. McDonald, O. Suleman, S. Williams, S. Howison (2005) 
`What shakes the FX Tree? Understanding Currency Dominance, Dependency and 
Dynamics', Proceedings of SPIE -- the International Society for Optical 
Engineering, Vol 5848: 86-99.

[20] Lillo, F {\&} R. Mantegna (2002) `Dynamics of a financial market index 
after a crisis,' Phys A, Vol 338: 125-134.

[21] Bikhchandani, S. {\&} S. Sharma (2000) $`$Herd behaviour in financial 
markets: A review$,$' International Monetary Fund, WP/00/48.

[22] Frankel, J. A {\&} A. K. Rose (1996) `Currency crashes in emerging 
markets: An empirical treatment', J. Int Econs, Vol 41 (3/4): 351-366

\begin{center}
\textbf{Table 1 Countries selected for Exchange Data}
\end{center}

\begin{center}
Currency and international quotation code
\end{center}

\begin{center}
(23/10/1995 - 31/12/2001)
\end{center}

\begin{table}[htbp]
\begin{center}
\begin{tabular}{|l|l|l|l|l|}
\hline
Currency& 
Code& 
& 
Currency& 
Code \\
\hline
Algerian Dinar & 
DZD& 
& 
Italian Lira & 
ITL \\
\hline
Australian Dollar & 
AUD& 
& 
Japanese Yen& 
JPY \\
\hline
Bangladeshi Taka & 
BDT& 
& 
Jamaican Dollar& 
JMD \\
\hline
Belgium Franc & 
BEF& 
& 
Kazakhstan Tenge& 
KZT \\
\hline
Bolivian Boliviano& 
BOB& 
& 
Kenyan Shilling& 
KES \\
\hline
British Pound& 
GBP& 
& 
Malaysian Ringgit& 
MYR \\
\hline
Brazilian Real& 
BRL& 
& 
Mexican Peso & 
MXN \\
\hline
Canadian Dollar& 
CAD& 
& 
New Zealand Dollar& 
NZD \\
\hline
Chilean Peso& 
CLP& 
& 
Pakistan Rupee& 
PKR \\
\hline
Colombian Peso & 
COP& 
& 
Papua New Guinea Kina& 
PGK \\
\hline
Czech Koruna& 
CZK& 
& 
Philippine Peso& 
PHP \\
\hline
Danish Krone& 
DKK& 
& 
Polish Zloty & 
PLN \\
\hline
Dutch Guilder& 
NLG& 
& 
Qatar Rial& 
QAR \\
\hline
Eqyptian Pound & 
EGP& 
& 
Russian Rouble& 
RUB \\
\hline
Fiji Dollar& 
FJD& 
& 
Saudi Arabian Riyal& 
SAR \\
\hline
Finnish Markka& 
FIM& 
& 
Singapore Dollar& 
SGD \\
\hline
French France& 
FRF& 
& 
South Korean Won& 
SKW \\
\hline
Ghanaian Cedi & 
GHC& 
& 
South African Rand& 
ZAR \\
\hline
German Deutsche Mark& 
DEM& 
& 
Swedish Krona& 
SEK \\
\hline
Indian Rupee & 
INR& 
& 
Swiss Franc& 
CHF \\
\hline
Indonesian Rupiah& 
IDR& 
& 
Taiwan Dollar& 
TWD \\
\hline
Iranian Rial& 
IRR& 
& 
Thai Baht& 
THB \\
\hline
\end{tabular}
\label{tab1}
\end{center}
\end{table}

\end{document}